
\magnification=1200
\hsize=13cm
\def\newline{\hfil\break}  
\def\proclaim#1#2{\medskip\noindent{\bf #1}\quad \begingroup #2}
\def\endproclaim{\endgroup\medskip}
\def\<{{<}} \def\>{{>}} 
\font\Bbb =msbm10
\def\Real{{\hbox{\Bbb R}}} 
\def\H{{\cal H}} \def\B{{\cal B}} \def\F{{\cal F}}

\def\parasign{\S}
\def\blacksquare{\vrule height 4pt width 3pt depth2pt}
\def\dsize{\displaystyle}
\abovedisplayskip=3pt plus 1pt minus 1pt
\belowdisplayskip=3pt plus 1pt minus 1pt

\font\brm=cmbx12  \def\tilbf{\lower 1.1 ex\hbox{\brm \char'176}}

\font\trm=cmr12  
\def\tilrm{\lower 1.1 ex\hbox{\trm \char'176}}

\topskip10pt plus40pt
\clubpenalty=30

\headline={\hfil}
\footline={\hss\tenrm\folio\hss}

{\bf 
\centerline{Finitary and Infinitary Mathematics,}
\centerline{the Possibility of Possibilities and the Definition of Probabilities.
\ 
\footnote{*}{\tenrm quant-ph/0306201, June 2003.}}
\medskip

\centerline{Matthew J. Donald}
\medskip

\centerline{The Cavendish Laboratory,  Madingley Road,}  

\centerline{Cambridge CB3 0HE,  Great Britain.}
\smallskip

\centerline{ e-mail: \quad matthew.donald@phy.cam.ac.uk}
\smallskip
{\bf \hfill web site:\quad  
{\catcode`\~=12 \catcode`\q=9
http://www.poco.phy.cam.ac.uk/q~mjd1014
}\hfill }}
\medskip

\noindent{\bf Abstract} \quad  Some relations between physics and
finitary and infinitary mathematics are explored in the context of a
many-minds interpretation of quantum theory.  The analogy between
 mathematical ``existence'' and physical ``existence'' is considered from
the point of view of philosophical idealism.  Some of the ways in which
infinitary mathematics arises in modern mathematical physics are
discussed.  Empirical science has led to the mathematics of quantum
theory.  This in turn can be taken to suggest a picture of reality
involving possible minds and the physical laws which determine their
probabilities.  In this picture, finitary and infinitary mathematics play
separate roles.  It is argued that mind, language, and finitary
mathematics have similar prerequisites, in that each depends on the
possibility of possibilities.  The infinite, on the other hand, can be
described but never experienced, and yet it seems that sets of
possibilities and the physical laws which define their probabilities can
be described most simply in terms of infinitary mathematics. 
\medskip

This is an extended version of a talk given to an audience of mixed
backgrounds in a philosophy of mathematics seminar series in Cambridge. 
My aim is to explore some ideas about the relation between mathematics
and reality in the context of the version of the many-minds interpretation
of quantum theory which I have been developing for many years.  I have
been led to my present radical views on reality as a result of taking the
mathematics of quantum theory all too seriously.  The justification of these
views requires that the structure of reality is governed by mathematical
law.   But what sort of mathematics is ultimately required?

Philosophers of mathematics have made many suggestions about the
nature of mathematics and almost all of them have some validity. 
Mathematics is useful if-thenism and a language game and a social practice
and a means of discovering eternal truths and reducible to set theory;
speaking of the existence of the set of real numbers as a completed infinite
totality is both useful and dangerous; numbers are structures which obey
the rules which would be obeyed by ratios of physical lengths in
a physical Euclidean geometry or the rules which would be obeyed by any
abstract structure given appropriate axioms.  Such a wide range of good
ideas seems to arise because there are so many different levels at which
mathematics is important.  Unfortunately,  philosophers of mathematics are
often tempted to be dogmatic and to claim that mathematics is nothing but
useful if-thenism, or nothing but a language game, or nothing but a social
practice, or nothing but set theory; or only certain types of construction are
to be allowed; or that numbers are nothing but sets, or nothing but ratios of
lengths.  Such dogmatic claims are rarely placed in the context of the sort of
thorough analysis of the nature of reality which should surely be required
to justify them, especially given the importance of mathematics to
physics.  The dogmatic claim I wish to investigate in this paper is the idea
that, ultimately, only finitary mathematics is indispensable.

The idea in the philosophy of mathematics, that a certain aspect of
mathematics is ``indispensable'', is, broadly, the idea that that aspect is
required for science and that therefore we ought to accept its reality. 
``Finitary mathematics'' is mathematics which can be expressed without
invoking infinite sets.  This will be left as a fairly vague notion here.  For
example, the fact that $\sqrt{2}$ is irrational is infinitary, but this does not
imply that all circumstances in which we refer to a ``number'' $r$ such that
$r^2 = 2$ involve infinitary mathematics.  The question of how this
vagueness might be eliminated will be ignored, because we shall focus on
the idea of ``ultimately indispensable'' and explore instead some questions
about the possibility of a serious and plausible complete scientific
understanding based on a finite ontology.

My analysis of quantum theory is one motivation for exploring these
questions.  Quantum theory has led me to a form of philosophical idealism. 
According to this, mind is the primary aspect of reality and each individual
mind is finite.  Nevertheless, infinitary mathematics still seems to be
needed for the definition of probabilities of individual mental states.  My
personal goals in the work leading up to this paper have been to try to
understand the relationship between the finitary and the infinitary in this
context and to consider whether the use of infinitary mathematics can be
avoided.  However, although it is often illuminating, and may ultimately
be necessary, to think about philosophical questions in a specific
framework, the details of my interpretation of quantum theory are
certainly not relevant to the present paper and many of the questions
raised seem to me to be significant in a range of different contexts.

For example, mathematical physics certainly does make free use of the
mathematics of the infinite; irrational numbers, Minkowski space, partial
differential equations, infinite dimensional Hilbert spaces, type III von
Neumann algebras, even the axiom of choice is sometimes invoked.  On the
other hand, even without going as far as idealism, we should perhaps find
the idea of a physical reality which is actually infinite somewhat
disconcerting, in as far as we ourselves are finite beings, unable, for
example, to make any measurements with unlimited accuracy.  Moreover,
because we are finite beings, any application we make of infinitary
mathematics can always be replaced by an application of finitary
mathematics with indiscernable consequences.  This is important in
mathematical physics, because it means that any significant consequence of
infinitary mathematics must reflect facts that can be expressed with
finitary mathematics.  Often this is conceptually straightforward; as in the
process of constructing a discrete version of a differential equation for
numerical analysis.  Sometimes, however, it is not immediately clear what
sort of finitary fact might be involved and finding such a fact may be a
valuable part of understanding the application of the mathematics. 

As a functional analyst, I do not take kindly to being told, even by myself,
that I should not talk about the Lebesgue integral or the axiom of choice. 
Thus, I shall not even begin to argue that infinitary mathematics is without
validity.  It is surely hard, for example, to doubt the truths of logical
implication, in particular in as far as to doubt such implications is to doubt
the means of expression which are required if doubt is to be meaningful. 
Our finite proofs give us adequate reason to believe that the axiom of
choice is equivalent to Tychonoff's theorem and to Zorn's lemma.  But,
although, as we shall see, it may be useful to assume that arbitrary
products of compact spaces are compact (Tychonoff's theorem), it does not
necessarily follow that the axiom of choice is a necessary property of real
infinite sets  or, indeed, that there are any real infinite sets; any more than
the fact that it is useful to know that the diagonal of a square of side $a$
has length $a \times \sqrt{2}$ means anything more than that any real
physical square has side $a$ and diagonal $d$ which satisfy $d^2 = 2 a^2$ to
a degree which depends on the extent to which the square approaches an
unobtainable perfection.  This line of argument makes it tempting to think
that the mathematics of the infinite might be simply a language game;
simply the study of proofs.  The ultimate conclusion of this would be
the idea that although finitary mathematics tells us necessary truths about
physical reality, there might be no such necessary truths involving
infinitary mathematics, because physical reality simply has no infinite
aspects.  This would not make, for example, theorems about the
independence of the continuum hypothesis any less interesting as
mathematics, but it would remove any remnant of the intuition that the
continuum hypothesis must be true or false because of the actual existence
of real physical continua.

It seems that the many good ideas in the philosophy of mathematics need
to be re-worked again and again as we probe at the different levels at
which mathematics is important.  There are many excellent surveys of
such ideas.  I have enjoyed learning from and agreeing and disagreeing
with those discussed in Chihara (1990), Hersh (1997), Burgess and Rosen
(1997), and Maddy (1997).  In particular, there is an interesting
discussion of whether infinitary mathematics is indispensable in
Maddy (1997, chapter II.6).  In this paper, I shall probe
mathematics at the level of a working mathematical physicist, and then at
the level of physical law, and finally at the level of individual mental
processes.  The result will be to emphasize the gulf between finitary and
infinitary mathematics and to relate that duality to dualities between
events and their probabilities, and between experience and description,
and even between realism and fantasy.

\proclaim{What is Mathematics About?}
\endproclaim

Whatever mathematics may be, it is constructed out of theorems.  In this
section and the next, therefore, I shall analyse some theorems in order to
illustrate some ideas about the nature of mathematics and to demonstrate
some of the quite subtle ways in which the mathematics of the infinite
does arise in modern mathematical physics. 

If asked to calculate 17 times 19 in your head, you might work it out
from 17 times 20 minus 17, or, remembering that $(n-1)(n+1) = n^2 - 1$,
you might realise that it was 18 squared minus 1, and then you might have
forgotten the value of 18 squared, and would need to think of 4 times 81. 
You certainly wouldn't think about successors of 0, let alone about sets of
sets of sets of the empty set.

\proclaim{Theorem One}
$$17 \times 19 = 17 \times 20 - 17 = 18^2 -1 = 4 \times 81 - 1 = 4
\times 80 + 4  - 1.$$
\endproclaim

This theorem illustrates the fact that is no single or canonical path to
mathematical truths.  It may be possible to codify the logical structure of
any theorem in set theoretic terms, but that is not the same as capturing
the theorem's meaning.  Theorem one can also be used to illustrate some of
the ways in which mathematical facts can be represented and confirmed by
physical processes.  For example, the theorem could be confirmed by
drawing, re-arranging, and counting dots on a page, or by using grains of
rice on a table, or by using a calculator.

\proclaim{Theorem Two}
$${\pi^2 \over 6} = \sum_{n=1}^\infty {1 \over n^2}.$$
\endproclaim

This beautiful result is frequently mentioned by philosophers of
mathematics.  It will be used here to provide further illustration of the
multiplicity of meanings which can be attached to any mathematical truth.
Indeed, there is multiplicity of meaning even in the symbol $\pi$ used in
the statement of the theorem.  This symbol could be glossed as $I =
\dsize{2\int_0^1 {dx\over \sqrt{1 - x^2}}}$, which could, in turn be
considered as a line integral measuring distance along a set of points $(x,
y) \in \Real^2$ such that $x^2 + y^2 = 1$.  Considered as a Riemann
integral, $I$ has similar character to the convergent sum on the right
hand side.  $\pi$ might alternatively be defined as the first strictly positive
zero of the function $\sin x$ defined as $\sum\limits_{n=0}^\infty (-1)^n
x^{2n+1}/(2n + 1)!$, or as $\dsize\left( \int_{-\infty}^\infty e^{- x^2} dx
\right)^2$, or as $\dsize 4(\sum_{n=0}^N {(-1)^n \over 2n + 1} +
(-1)^{N+1} \int_0^1 {x^{2(N+1)} \over 1 + x^2} dx)$. 

There is also a multiplicity in the techniques by which the theorem can be
proved.  A quantum mechanic would expand the function $\varphi(x) = x$ on
$L^2[0,1]$ in the orthonormal basis $(\sqrt{2}\sin n \pi x)_{n\geq 1}$, and
obtain
$\varphi = \sum\limits_{n=1}^\infty {\sqrt{2} \over n \pi} (-1)^{n+1}
\varphi_n$, and hence
${1 \over 3} = ||\varphi||^2 = \dsize\sum\limits_{n=1}^\infty {2 \over n^2
\pi^2}$.

A pure mathematician might prefer a proof using uniform convergence
rather than convergence in $L^2$.  This can be done by using Fourier series
techniques to prove that 
${{1\over 4}(x-\pi)^2 = {1\over12}\pi^2} + 
\dsize{\sum_{n=1}^\infty {\cos n x \over n^2}}$ on the closed interval $[0, 2\pi]$. 

A more significantly different proof uses complex analysis and
Liouville's theorem to show that $\dsize{\pi^2 \over \sin^2 \pi z} =
\sum_{n\in\Bbb{Z}}{1\over (z - n)^2} \sim {1\over z^2} +
{\pi^2\over 3} + O(z^2)$.

Another proof uses Cauchy's theorem and involves integrating the function
\newline $\dsize{\log (1 - z) \over z}$ around the half circle 
$\{ z = r e^{i\theta}: 0 \leq r \leq 1, 0 \leq \theta \leq \pi\}$ indented at $z
= 1$.  This proof can be modified and rewritten to use nothing beyond
advanced calculus on real functions.

Just as the different methods for theorem one use different numbers, these
proofs use different theories, different functional relations, different
integrals, and different expansions.  They all, however, ultimately require
the idea of convergence as does the very statement of the theorem.  The
truth of theorem two can thus be seen as a fact of infinitary mathematics. 
However, it is also a succession of facts about approximations in finitary
mathematics.  For example, using the last of the proposed definitions of
$\pi$, the theorem says that for any rational number $q > 0$, there is a
natural number $N_0$ (which can be explicitly estimated) such that $N
\geq N_0$ implies
$$|{8 \over 3}(1 - {1\over 3} + {1 \over 5} - \dots + {(-1)^N \over 2N +
1})^2 - (1 + {1\over 4} + {1 \over 9} + \dots + {1 \over N^2})| < q.
\eqno(1)$$

Facts of this sort can be confirmed by physical methods.  Indeed, any
inequality involving sums and differences of rational numbers can be
converted into an inequality involving sums and differences of integers.
It is possible to imagine this being checked directly using piles of pebbles. 
More practically, using a calculator to check the result will involve the
performance of a long series of elementary electronic operations.  Each of
these operations is a physical model of a mathematical process on finite
digit binary numbers. 

Of course, theorem two might also be taken to express facts about physical
circles; not only with $\pi$ characterized as the ratio of circumference to
diameter, but also, for example, with $\pi$ characterized by the Buffon
needle problem as equal to $2L/p D$ where $p$ is the probability that a
needle of length $L$ hits a line when it is dropped in a suitably random
way onto a floor marked with parallel lines separated by uniform distance
$D$.  General relativity tells us that, in general, such facts are actually false,
but there are a wide variety of possible physical processes which could
provide confirmations that with such characterizations of  $\pi$ the
theorem is approximately true.

The multiplicity of ideas which can be expressed through theorems one and
two suggests to me that, first and foremost, mathematics is about
mathematics.  In particular, I would apply this to the idea of mathematical
existence.  When we say, for example, that $\dsize{\pi^2 \over \sin^2 \pi
z}$ \vadjust{\kern1pt} has a second order pole at $z = 13$, we are using talk
about the ``existence'' of a pole as a shortcut for talk about the
consequences of that ``existence''; for example, that $\dsize{\pi^2 \over
\sin^2 \pi z} \sim {1 \over (z - 13)^2}$ for $z \sim 13$ or that if we draw a
diagram of the singularities, we had better, if our diagram goes that far,
put a dot or a cross, or whatever symbol we want to use, at co-ordinates
$(13, 0)$. 

Idealism applies a similar idea of existence in scare quotes to physical
objects.  An idealist who says that a carelessly kicked stone can break a
window, is using talk about ``stones'' as a shortcut for talk about the
apparent consequences of the ``existence'' of ``stones''  -- for example, the
feeling of the ``kick'' and the sound of ``breaking glass'' and the subsequent
fear or embarassment.  An idealist turns talk about existence into talk
about consequences and reduces all consequences to mental terms.  In
mathematics, it is sufficient to reduce all talk about consequences to talk
about mathematics, and hence to talk about what mathematicians do or
might do, and hence to talk about what they see themselves as ``doing'' or
possibly ``doing''.

  To justify idealism, it is necessary to explain the
coherence of the world of appearance.  In my many-minds interpretation
of quantum theory, the explanation will involve the idea that minds are
only likely to exist in as far as it is likely for rich patterns of information to
develop, and that this is only possible in circumstances in which the stable
repetition and development of patterns is likely.

The coherence of mathematics also needs to be explained.  Explanations
in terms of the realistic existence of mathematical objects may be
rejected, on the grounds that it is neither clear how such existence could be
constituted, nor how it could explain our knowledge.  The fact that our
understanding of the idea of $\pi$, for example, has so many different
aspects should be seen not as a reflection of the existence of some perfect
Platonic form, but rather of the importance of ideas about ``circles'' to
beings who try to make sense of our sort of reality.  In other words, our
interest in rules which would govern ideal Euclidean circles, if they existed,
does not imply that such circles do exist, but merely that it is possible and
useful for us to make the abstractions which are needed to allow us to talk
about them or to ``posit'' them (Quine 1951, \parasign VI).  In
conventional, non-idealist, terms, this would be to say that imperfect
circles do exist and that it is by trying to understand and simplify rules for
imperfect circles that we come to discover rules for perfect circles. 
According to my interpretation of quantum theory, imperfect circles also
do not exist.  Instead, we exist as rich structured finite meaningful
stochastic patterns which have high probability of short term continuation. 
It seems plausible that the simplest way in which that sort of existence is
possible is if the patterns can give themselves some sort of geometric
meaning.  Ultimately, abstractions about ideal Euclidean circles can be
made because collections of finitary facts like (1) are true. 
The possibility of and the motivation for discovering such truths are part of
the structure of human reality at the deepest level.

\proclaim{Taming the Infinite in Mathematical Physics.}
\endproclaim

In this section, we shall consider three examples of the relation between
mathematical physics and the infinite.  Theorem three introduces a
situation in which infinitary mathematics provides a model which is both
convincing and clearly unrealistic.  Theorem four seems, at first sight, to be
both  significant and necessarily infinitary.  It therefore provides a
challenging example of the question of whether any application of
infinitary concepts can always be expressable in finitary terms.  The final
situation exemplifies the use of the axiom of choice in mathematical physics.

\proclaim{Theorem Three}
There are no phase transitions in finite systems.
\endproclaim

One of the central problems of mathematical physics is to gain a
theoretical understanding of the phases of matter -- such as ice, water, and
steam -- starting from a quantum mechanical description of the electrons
and nucleons involved.  Much progress has been made by the analysis of
model systems (Ruelle 1969, Bratelli and Robinson 1981, Krieger, 1996). 
In such systems, a phase transition is defined as a singularity of a
thermodynamic function, and there are models which do exhibit phase
transitions of this nature which have properties which appear to reflect
the properties of observed physical phase transitions.   However, there
are no such singularities in systems which model finite numbers of
molecules.  This means that, however realistic our models become, those
phase transitions which satisfy the definition will only be a caricature
of our observations.  This is a strange situation in which our models
appear to become more realistic, in that they describe changes of phase,
only when they become less realistic, by describing infinitely many
molecules.  

There is no fundamental problem here, but although this is a case in which
at least one infinitary aspect of the mathematics used is clearly not
indispensable in that, of course, we cannot take the accuracy of this
aspect of infinite models of physical systems as a demonstration that we
are wrong to believe that Avogadro's number (a measure of the number of
molecules in objects weighing, very roughly, a hundred grams) is actually
finite, nevertheless it would be an over-simplification just to say that the
mathematical definition of a phase transition is incorrect.  In fact, we can
never wait long enough for a real macroscopic system to reach a true
equilibrium state; and anyway, for low temperatures, we could no more see
such a state than we could see a cat being both alive and dead.  On the
other hand, the ergodic equilibrium states of infinite systems are actually
excellent models of the observed quasi-equilibrium states of real (finite)
macroscopic systems.

\proclaim{Theorem Four}
There are no pure normal states on a Type III von Neumann algebra.
\endproclaim

This is a theorem which I believe may be of fundamental importance for
the foundations of quantum theory in that it can be interpreted as
saying that quantum theory should not ultimately be seen as a theory about
wavefunctions.  Wavefunctions are pure normal states on Type I von
Neumann algebras, but there are good arguments for believing that local
systems in quantum field theory should be described by Type III algebras
instead (Haag 1992, \S V.6).  Quantum field theory is the most
complete version of quantum theory available, and observers of course are
localized systems.  Type III von Neumann algebras, however, require
infinite dimensional systems and this raises the question of how any
useful aspects of this theorem might be expressed in finitary terms. 
Answering this  question is not only a matter of soothing anxieties about
the infinite but also of revealing the physics behind what might
otherwise seem just  mathematical name-dropping.

A caricature of the theorem is possible in commutative quantum theory;
a subject which is sometimes referred to as ``probability theory''.  On a
finite set $\{1, \dots, n\}$, a ``state'' is simply a probability
distribution -- a sequence $(p_i)_{i=1}^n$ satisfying $0 \leq p_i \leq 1$ for
$i = 1, \dots, n$, and $\sum_{i=1}^n p_i = 1$.  A state $(p_i)_{i=1}^n$ is
defined to be ``pure'' if and only if there is no decomposition into distinct
states $(u_i)_{i=1}^n$ and $(v_i)_{i=1}^n$ such that, for $0 < x < 1$,
$p_i = x u_i + (1-x) v_i$.  It is quite easy to see that pure states exist and
are exactly those probability distributions such that $p_i = 1$ for some $i$.  

Now consider probabilities on an infinite set.  On the real line $\Real$,
various types of probability distributions are possible, but the ones which
correspond to normal states in quantum theory are the distributions
defined by bounded measurable functions $p \in L^\infty(\Real, dx)$ such
that $0 \leq p(x)$ almost everywhere and such that the probability of $A
\subset \Real$ is given by $P(A) = \int_A p(x) dx$.  This kind of state can
never be pure, because it is always possible to find disjoint sets $A_1, A_2
\subset \Real$ such that $P(A_1) = P(A_2) =  {1\over 2}$ and then $p(x) = 
{1\over 2} (2 I_{A_1}(x) p(x) + 2I_{A_2}(x) p(x))$ where $I_{A}$ is the
characteristic function of the set $A$.  Those familar with the concepts
in theorem four will prove it using a related idea -- a pure normal state
has a minimal support projection, but type III algebras have no
minimal projections. 

A pure state in quantum mechanics is a state which provides maximal
information.  In commutative quantum mechanics, pure states provide
complete information (certainty).  The non-existence of normal pure
states in the infinite realm means that complete information is completely
unobtainable.  This is a good model of a situation in which it is difficult to
obtain complete information.  With classical mechanics on continuous space,
it would have been absurd to try to find out by measurement whether the
ratio of distances between two pairs of particles was a rational number.  On
the other hand, given a reasonable model of measurement, exact
measurements on a discrete lattice may be expected to become increasing
hard as the lattice spacing decreases.  This means that from the point of
view of a classical physicist who accepts that space is either continuous, or
discrete at an imperceptibly short length scale, a theory, like quantum
mechanics, in which the distance between a pair of particles cannot have
an exact value is not obviously false.

In the application of theorem four, I would argue that we are in the
situation that either wavefunctions do not exist for real localized
systems, or that we cannot tell whether a quantum state on a real system
is pure or mixed.  In either case, a theory of quantum mechanics based on
mixed states rather than wavefunctions cannot be obviously false.

Infinitary mathematics plays an important role in this sort of analysis. 
It introduces a conceptual issue in an extreme form (``pure states are
impossible'') and it also presents us with the mathematical challenge of
working out how that issue can be expressed in approximations to the
extreme situation.  For example, if theorem four is physically significant,
then it should tell us something physically significant about the nature of
finite-dimensional quantum systems of large dimension.  One possible
statement of such a fact would be that the quantum entropy ($S$ -- a
measure of the purity of a state) is hard to control in Hilbert spaces of large
dimension ($D$).  For example, the following proposition, which follows
straightforwardly from the concavity of
$S$ and the existence of a state $\tau$ with $S(\tau) = \log D$, shows that in a
Hilbert space of sufficiently large dimension, there is a state of large
entropy close to any given state. 

\proclaim{Proposition}{\sl}  Given $\varepsilon > 0$ and $M > 0$, there
exists $D$ such that, for any state $\rho$ on a Hilbert space $\H$ of
dimension at least $D$, there exists a state $\sigma$ on $\H$ such that
$||\rho - \sigma|| < \varepsilon$ and $S(\sigma) > \log M$.
\endproclaim

Note that this proposition is by no means entirely free of infinitary
mathematics and neither is its interpretation entirely straightforward.  The
aim here, however, is merely to illustrate the fact that it would be a
mistake in physics not to try to understand how any given invocation of
infinitary mathematics could be expressed in finitary terms.

\proclaim{Theorem Five}
A product of compact spaces is compact.
\endproclaim

This is Tychonoff's theorem, which, as I have mentioned, is equivalent to
the axiom of choice.  The various forms of the axiom of choice are useful in
functional analysis because they can be used to prove that various
``objects'' ``exist''.  For example, the Hahn-Banach theorem, which is proved
using Zorn's lemma, shows the ``existence'' of extensions of linear
functionals.  Tychonoff's theorem can be used to demonstrate the
``existence'' of limits of generalized sequences.  This is similar to the way in
which the supposed completeness of the real numbers can be used to
demonstrate the ``existence'' of a positive number $r$ satisfying $r^2 = 2$. 

The limits, however, can be considerably less innocuous than $\sqrt{2}$. 
For example, theorem four referred to ``normal'' states on a von Neumann
algebra.  On a finite-dimensional algebra, there are no other states, but in
infinite dimensions, Tychonoff's theorem implies the w$^*$-compactness of
state space considered as a closed subset of the dual of the algebra.  This
means, for example, that if $(\psi_n)_{n=1}^\infty$ is an orthonormal basis
for an infinite-dimensional Hilbert space $\H$, then the sequence
$(|\psi_n\>\<\psi_n|)_{n=1}^\infty$ of density matrices has a convergent
subnet.  In other words, there is a positive linear functional $\rho$ on the
space $\B(\H)$ of bounded operators on $\H$ such that, for all $A \in
\B(\H)$, $\rho(A)$ is a limit point of the sequence
$(\<\psi_n|A|\psi_n\>)_{n=1}^\infty$.  In particular, $\rho(1) = 1$ and yet
$\rho$ is singular in the sense $\rho(|\psi_n\>\<\psi_n|) = 0$ for all $n$. 
This shows that $\rho$ is not $\sigma$-additive.   In the language of
probability theory, $\rho$ defines a probability $P$ on the positive integers
which is additive but not $\sigma$-additive, and which vanishes on finite
sets.  However, $P$ is far from completely defined by this statement.  Given
any infinite subset $S$ of the positive integers, we can, for example, choose
$P$ to satisfy $P(S) = 1$.  Then, given any infinite subset $S' \subset S$, we
can choose $P(S') = 1$.  Infinitely many distinct choices are required to
specify $P$ (or $\rho$).  To provide a satisfactory mathematical theory of
such states, and in particular to define pure singular states, it is necessary
to use a form of the axiom of choice stronger than the axiom of dependent
choice (Halvorson 2001).

It would be good to rule all non-normal states out of consideration in
physical applications.  This can be certainly done, as I have implied in
discussing theorem four.  Indeed, physical states can be assumed to have
all sorts of physically-required properties in addition to normality, such as
finite energy.  Nevertheless, mathematical considerations can lead to the
use of non-normal states in theorems of physical relevance.  For example,
in the development of a functional on pairs of quantum states which I
believe to be relevant to the calculation of quantum probabilities,  I
invoked the w$^*$-compactness of the space of states in order to give a
simple definition in terms of suprema (Donald 1986, 1992).  To make such a
definition acceptable, it has to be accompanied by theorems showing that
whenever non-normal states might arise, they have no necessary place in
the physics (Donald 1986, Axiom IV, Donald 1992, Property K).

\proclaim{Infinite Approximately-Homogeneous Cosmologies.}
\endproclaim

There are many different aspects of physical reality which might be
infinite.  Some of these infinities could be more problematic than others. 
In particular, one idea which is often thought plausible in cosmology is that
of a universe which is spatially or temporally infinite and which is
approximately homogeneous.    The sense of approximately homogeneous
might quite reasonably be taken to be that there are infinitely many
spacetime regions of the magnitude of our visible universe in which the
same laws of physics apply and in which roughly the initial conditions of
our big bang applied.  In that case, we can apply the following theorems.

\proclaim{Theorem Six}
Let $I$ be an infinite set and $F$ be finite.  Let $f : I \rightarrow F$.  Then
there are infinitely many elements of $I$ which have the same image
under $f$.
\endproclaim

The theorem shows that if the universe contains infinitely many stars or
planets or lifeforms or intelligent lifeforms or mathematical proofs, then
there are infinitely many stars or planets or lifeforms or intelligent
lifeforms or mathematical proofs which are arbitrarily similar by any finite
measure.  Given homogeneous laws of physics, a suitable finite measure
might specify the approximate relative positions of a finite number of
isotopically-identified species of individual atom over a finite sequence of
times.

\proclaim{Theorem Seven}
Let $(A_n)_{n=1}^\infty$ be an infinite sequence of independent events for
some probability $P$ and suppose that, for some $\delta > 0$ and all $n$,
$P(A_n) \geq \delta$.  Then, with probability one, infinitely many of the
$A_n$ occur.
\endproclaim

This theorem is a version of a Borel-Cantelli lemma.  Applied to an
approximately homogeneous infinite cosmology, it says that if it is possible
for a particular star or lifeform or mathematical proof to exist in our visible
universe, then, given the observed indeterminism of quantum processes, it
is inevitable that infinitely many such stars or lifeforms or mathematical
proofs exist which are arbitrarily similar by any finite measure.  This is a
sufficiently striking idea that it seems worth giving a proof of theorem
seven.
\medskip

Let $X_N$ be the event that no $A$ after $A_N$ happens.
$X_N$ is less likely than that, for any $M$, $A_{N+1}$ and $A_{N+2}$
and \dots and $A_{N+M}$ don't happen, so that, using independence,
$P(X_N) < (1 - \delta)^M$  for all $M$, and hence $P(X_N) = 0$.

Let $X$ be the event that only finitely many of the $A_n$ happen. 

If $X$ happens then, for some $N$,  $X_N$ happens, and so
$P(X)$ is less than or equal to $P(X_1) + P(X_2) + P(X_3) + \dots$

Thus $P(X) = 0$, and theorem 7 is proved. \hfill $\blacksquare$
\medskip

The first part of this proof suggests that if you have to throw double six to
get out of jail, then you will get out if you keep throwing long enough. 
Of course, this ignores the possibility that you might die first.  The second
part of the proof is the idea that the union of countably many events
of probability $0$ has probability $0$.

It is sometimes argued that probability in a many-worlds interpretation of
quantum theory is suspect in as far as such a theory is fundamentally
deterministic with all possible future events actually happening in one
world or another.  Theorem seven shows that an approximately
homogeneous infinite cosmology is somewhat similar.  In neither situation
is it entirely reasonable to say, when faced with a pair of distinct possible
outcomes, ``Either A or B, but not both is going to happen'', because in both
situations, nothing observed about the prior circumstances distinguishes
between the apparently inevitable futures in which A and B occur.  This
apparent inevitability may however only be apparent in an idealist
many-minds interpretation, such as my own, in which physical laws
provide probabilities for minds to exist but it need not necessarily be the
case that all possible consciousnesses are experienced. 

\proclaim{Empirical Science and Infinitary Mathematics.}
\endproclaim

Even if our interactions with reality are entirely finite, and even
if all of our observations can be explained in terms of finite models, it does
not seem particularly plausible that they should be so explained. 
Euclidean geometry, Newtonian mechanics, thermodynamics, statistical
mechanics, electromagnetism, special and general relativity, the
Schr\"odinger equation, quantum electrodynamics, the standard model; the
best empirically-validated theories, which have been used to explain and
develop the technological achievements of our civilization, all depend on
infinitary mathematics.  Of course no theory comes with a guarantee of
truth, and in fact, there are empirical gaps or failures in every known
theory.  In particular, we have yet to discover any coherent ``theory of
everything'' which convincingly combines general relativity with quantum
theory.  It is just about conceivable that there might be a coherent and
convincing theory of everything which depended entirely on finitary
mathematics.  But this seems pretty unlikely.  Successful extensions and
revisions of previous theories have always involved trying to work
empirically derived concepts into simple and beautiful mathematical
structures.  Finitary mathematics however is rarely as simple, let alone as
beautiful, as infinitary mathematics.  It is easier, at least given appropriate
training, to think of and to use $\sqrt{2}$ as a ``number'' rather than
constantly to refer to any of the many ways of approximating that
irrational by rationals.  The continuous differential equations of
mathematical physics are far more beautiful than the machine and
code-dependent approximations to those equations which are
used in the computation of numerical solutions.  When cosmology
turned away from infinite flat Euclidean geometry, it became necessary
to wonder whether the universe might have a radius, and to ask how it
might develop in time.  If spacetime were actually a finite lattice of points,
then questions about how many points there might be to a metre would
arise.  Is it appropriate to raise such questions without empirical input, just
because of philosophical scruples about the mathematics of the infinite?

The crucial motivation for the application of infinitary mathematics in
theoretical physics is the empirical success of theories developed to satisfy
human taste in mathematical simplicity.  What seems strange about this is
that this simplicity is a simplicity of external description, which is not
necessarily the same as an ontological simplicity.  It is also not the same as
a simplicity of total description.

\proclaim{Theorem Eight.}
Let $I$ be an infinite set without distinguished elements.  Then an infinite
amount of information is required to identify any element of the set.
\endproclaim

This theorem holds because if any point in $I$ could be identified by a finite
description, then the finitely many points which could be identified with
the shortest description in a  given language would thereby be
distinguished.  It might be relevant to the problem of specifying an
individual star or lifeform or mathematical proof in an infinite
approximately-homogeneous cosmology.

\proclaim {Mathematics, Madness, and Quantum Mechanics.}
\endproclaim

Quantum mechanics provides some important examples of the power of
mathematical aesthetics to lead us towards ideas about the nature of
reality, which, although compatible with empirical observation, are so far
from common sense that they might be thought insane.

A lesser insanity, associated with the names of Einstein, Podolsky, Rosen,
Bell, and Aspect, comes from taking too seriously a small part of the
formalism of quantum theory and combining it with relativity theory 
(Bell 1987).  This leads to a belief in spooky and mysterious
connections between spatially distant points.  For example, there are
apparently situations in which one experimenter ``Alice'' can find out
which of two genuinely random events a distant colleague ``Bob'' will
observe, simultaneously with his observation, but without there being
anything in Bob's laboratory which predetermines his result, or any
message of any kind passed between the laboratories between Alice's
observation and Bob's.  This insanity unfortunately is essentially
unavoidable as its consequences have been convincingly and directly
demonstrated experimentally.

A greater insanity, from which I myself suffer in a particularly acute
form, goes back to the idea of the Schr\"odinger cat experiment
(Schr\"odinger 1935).  What this thought experiment indicates is
that, if the mathematical formalism of quantum theory, in the specific form
of the Schr\"odinger equation, gives an accurate picture of the behaviour of
ordinary macroscopic objects, then reality has to be pretty weird.

This weirdness was first taken at face value by Hugh Everett III
(1957), who, in the development of the many-worlds interpretation,
argued that the mathematics of quantum theory suggests that all possible
future possibilities, even macroscopically distinct possibilities like whether
a cat lives or dies, actually happen, but that being limited physical systems
ourselves, we can only see individual possibilities. 

Everett's work largely leaves open the questions of characterizing
individual possibilities and our limited natures.  My own approach to these
questions has involved the analysis of minds as finite systems processing
finite information.  As minds, we appear to live inside physical reality,
our brains apparently being direct physical representations of the
structure of our minds.  But, if minds are fundamental, then what we think
of as ``physical reality'' is just a mental representation.  What is ``external''
to mind, nevertheless, is the physical law which determines the
probabilities of the possible futures of a given mental structure.  

Although this picture is radical, it is, in my opinion, both logically consistent
and consistent with empirical evidence.  Moreover, it has considerable
explanatory power.  For example, mind is placed at the heart of reality,
rather than being just an embarrasment as it would seem to be for
materialists.  Because time becomes an aspect of our individual structure
as observers, the idea of a fixed observer-independent background
spacetime becomes unnecessary.  This may be useful in the development of
quantum gravity theories.  The problem of ``fine-tuning'' of physical
constants can be resolved by noting that if all information is mental, then
``constants'' have to be determined by observation, are fixed only to the
extent to which they have been observed, can only be observed if they
have values which make observation possible, and are only likely to be
observed if they have values which make observation likely.  The
situations in which one experimenter ``Alice'' can apparently find out
which of two random events a distant colleague ``Bob'' will observe before
his observation, but without there being anything in Bob's laboratory which
predetermines his result, or any message of any kind passed between the
laboratories between Alice's observation and Bob's, can be explained by
noting that Alice cannot confirm that Bob's observations satify her
predictions until the experimenters compare notes, and, at that stage,
effectively-local physical laws can ensure that Alice's observations of Bob's
results are compatible with the earlier observations from which she
deduced her predictions.  This explanation depends on Alice and Bob
having equally fundamental but separate mental lives. 

\proclaim{Dualities and Infinitary Mathematics in a Many-Minds
Interpretation of Quantum Theory.}
\endproclaim

For the purposes of the present discussion, the most important aspect of my
interpretation of quantum theory is the requirement that individual mental
structures be finite.  This does seem to be plausible.  In particular, a
fundamental discreteness does seem to be built into the actual operation of
the human brain, at the level, for example, of local neural firing.  It is also
possible to make at least some sense of the stochastic nature of observed
reality if that reality is like a finite game of chance.  It seems to me far
harder to understand what it would be like to be an observer if throws
in the game of life allowed an infinite number of distinguishable
possibilities each with infinitesimal probability.  In such a situation, finite
probabilities only exist for infinite sets of possibilities.  Supposing that an
observer cannot be aware of an infinite amount of information, it would
seem not unreasonable to attempt to identify an observer with a
finitely-definable equivalence class of observationally-indistinguishable
possibilities.

As well as finite mental structures and discrete stochastic events, the
theory also depends on physical laws.  My theory is a form of idealism, in
which the physical laws and initial conditions have no role or reality
beyond that of providing probabilities for mental histories.  These
probabilities are defined through the construction of abstract sets of
possible physical manifestations for each mental history.  A mental
structure is given a precise abstract definition and a ``possible physical
manifestation'' is one of the ways in which the laws of physics can permit
the development of something characterized as having that abstract
structure.  In the most complete current form of the theory
(Donald 1999), these physical laws are taken to be the laws of
special relativistic quantum field theory and the characterization allows for
uncountably many possible manifestations for any given mental structure. 
In this way, the theory manages to push to a higher level of abstraction the
problem of giving a precise characterization of an observable event; one of
the most difficult problems in the interpretation of quantum theory.  
Infinitary mathematics is invoked by the bucketful in the technical
appendix of Donald 1999, but that infinitary mathematics is
invoked entirely in the process of defining probabilities for finite mental
structures.

The gulf in a duality between experienced events and their probabilities is
much wider than that in the conventional duality between mental
experiences and the physical brains which are supposed to be possessed by
those experiences.  It is surely significant that infinitary and finitary
mathematics lie on either side of this gulf, even if the eradication of the
infinite remains incomplete.  Physicists are familiar with a duality between
initial conditions and dynamical equations.  In classical physics at least, the
dynamical equations are supposed to be simple and beautiful, while the
initial conditions carry, once and for all, all the assumed infinite
complexities of the assumed actual physical world.  The duality I am
proposing makes a considerable alteration to this picture.  The initial
conditions are replaced an entirely homogeneous initial state gradually and
stochastically embellished by the finite developing complexities of an
individual's actual observations up to that individual's current present. 
This, in fact, provides an explanation of the approximate homogeneity of
our observed cosmos.  The dynamical equations remain simple and
beautiful, but infinitary mathematics is required in the analysis of the ways
in which those complexities can arise and in the resulting actions of the
equations.

Yet another related duality is that between ``meaning'' and ``structure'' or
between ``readers'' and ``texts''.  In a fascinating recent book,
Tasi\'c (2001) explains how versions of this duality have
frequently appeared in the historical conversation between philosophy and
mathematics.  My work could be described, if I may be so bold, as
``scientific'' or ``ontological'' idealism rather than as  ``epistemological''
idealism.  This is to say that, in order to understand what physical theory
and empirical observation seem to be telling us about the nature of reality,
I have been led to propose a picture in which physical appearances are
constructs of fundamental mental structures. My primary concern has been
to use theory and observation to discover what form those structures
might take and how they might develop, rather than to try to understand
the process by which a given structure gives itself meaning or the
reliability of the knowledge obtained in such a process.  I am a realist
about reality, but the form of the fundamental structures of reality is not
immediately obvious to us, and our best hope of discovering that form lies
in the scientific circle of experiment, interpretation, theory, aesthetics,
mathematics, and consistency.

There are two ideas through which it might yet be possible to avoid the
necessity of infinitary mathematics in the context of my interpretation, or
an extension of my interpretation, of quantum theory.  Neither idea,
however, seems entirely satisfactory.

The first idea would avoid infinitary mathematics through the assumption
that the ultimate correct theory of everything was an entirely finitary
theory.  It is and always will be possible to make such an assumption, but it
is cheap for those not involved in the awesome difficulties of theories of
fundamental interactions to presume to say how those theories should turn
out.  Moreover, the original starting point which eventually led me to this
paper, lay not in a problem with infinitary mathematics as such, but rather
in the gulf in complexity between the resources required to simulate a
mental structure in my theory, and those required to simulate the
probabilities for the development of that structure.  With a finitary
fundamental theory, that gulf will still be wide, even if it is no longer
infinitely wide.

Given that in my proposals, infinitary mathematics is only required for
the definition of finite sequences of probabilities, the second idea for
avoiding the reality of infinitary mathematics would involve the fact that
an individual finite sequence of events cannot precisely determine the
probabilities of events making up that sequence.  In this scenario, the gulf
in resources required for a simulation would be narrowed by simulating
probabilities only to a precision which would make individual structures
likely to predict the correct theoretical probabilities.  In Donald
(2002), I argue that probabilities in a many-minds version of quantum
theory need to be precisely defined in order to make sense of the idea of
probabilities as theoretically-defined propensities.  If the probabilities are
not precisely defined, then the theory is incomplete.  Nevertheless, events
either happen or do not happen.  The only empirical justification for
accepting a probabilistic theory is that the events which have happened
are typical events according to that theory.  But a ``typical event'' is just an
event which belongs to some class of significant sets which have high
probability.

\proclaim{Theorem Nine}  Let $(A_n)_{n=1}^N$ be a finite sequence of
events generating a $\sigma$-algebra $\F$ in a probability space $(X, \F, P)$. 
Then, given any $\delta > 0$, there exists an integer $N$ and a probability $Q$,
such that $Q(A) \in \{ {M \over N}: M = 0, 1, \dots, N \}$ and  $|Q(A) - P(A)|
< \delta$ for all $A \in \F$.
\endproclaim

\proclaim{example}  Suppose that a single sequence of $N$ heads and tails
is produced by a sequence of independent events with the
probability of heads at each stage being $p = e^{-1}$.  

Compare this with a single sequence of $N$ heads and tails
produced by a sequence of independent events with the probability of
heads at stage $n$ being  $q_n =
\sum_{k=0}^n {(-1)^k \over k!}$.

For the first process, the expected number of heads is
$E^1(H) = e^{-1} N$ with a variance of $(e^{-1}-e^{-2})N$.
For the second process, the expected number of heads is
$E^2(H) = \sum_{n=1}^N \sum_{k=0}^n {(-1)^k \over k!}$.
$$E^1(H) - E^2(H) = \sum_{n=1}^N {(-1)^n (n - 1) \over n!} + N \sum_{n=N
+1}^\infty {(-1)^n \over n!}.$$
$$|E^1(H) - E^2(H)| \leq \sum_{n=1}^\infty { n \over n!} = e.$$

No possible statistical test can give a plausible argument to distinguish
which method of generation has been used to produce a single such
sequence, whatever the magnitude of $N$. 
\endproclaim

Ultimately, the generation of probabilistic events seems to be an entirely
mysterious aspect of reality.  Asking about it is like asking about where the
initial conditions come from in a deterministic theory, or like asking
why there is something rather than nothing.  However the question of the
extent to which infinitary mathematics is necessary in the generation
of such events is somewhat more approachable.  The intuition that I am
working with here is a picture of reality as being furnished with a random
number generator which provides, or has provided, a sequence of random
events of fixed probability -- say a sequence of bits of probability
$ {1\over 2}$.  Theorem nine indicates that each of the random events which make
up an individual's experiences could then be determined by a
finite number of those bits which could provide a choice indistinguishable
to that individual from the ideal probability generated by the infinitary
mathematics.  In this way, even the probabilistic generation of an
individual's experiences might only involve a finite complexity directly
reflecting the complexity of the experiences.  

Nevertheless, although it might theoretically be possible in this way to
avoid any actual infinitary mathematics in the construction of our reality,
it would be completely unscientific to ignore the manifest simplicity of the
physical laws which seem to govern that reality.  Indeed, surely simplicity
must underlie any explanation for the monotonous regularities of the
cosmos.  Galaxies, and lives, and suffering seem cheap in the economics of
creation; outcomes and purpose seem expensive.  We appear to ride the
crest of an evolutionary wave which formed us merely because our
ancestors happened to be able to survive.  My interpretation of quantum
theory amounts to the definition of a stochastic process on abstract
patterns of information.  Despite the complexity of this definition, it seems
to me far from inconceivable that among all the possible ways in which
stochastic processes can be defined on such patterns, the postulated
definition is actually as simple as any comparable set of rules which make
likely rich and meaningful patterns of information.   ``Simplicity'' in this
context, however, is the simplicity of infinitary mathematics, which is a
simplicity of external description rather than a simplicity of construction. 
Is my desire to give an explanation compatible with the latter as well as the
former merely yet another anthropocentric mistake about the economics of
creation?

These remarks seem excessively speculative.  Perhaps any attempt at a
thorough analysis of the nature of reality is bound to end in speculation
and mystery, but at least speculation is better than dogma.  In the next
section, we shall leave the mysteries of the generation and meaning of
probabilities and return to our world of finite experiences to consider
similarities between the fundamental structures of mind, language, and
finitary mathematics.

\proclaim{The Possibility of Possibilities.}
\endproclaim

Mind, language, and finitary mathematics are mutually fundamental and
each depends on the possibility of possibilities.  Language represents
meaning through finite structured patterns.  Mathematics is the study of
structured patterns.  Mind is a finite structured pattern which discovers its
own meaning through being the history of its development.  Structures
carry information only because they could be other than they are.  Words
are only interesting if something else might have been said.  Mind is
awareness of external reality as external.  It is awareness of possibilities of
pain and of comfort, of hunger and of satisfaction, of interest and of
boredom.  My characterization of an ``observer'' as a finite
information-processing structure requires an analysis of such a structure
as a pattern of repetitions and denials of abstractions of elementary
localized events.  For such a pattern to be meaningful, it must recognise
itself as a carrier of information in the same way that the meaning of a
language is recognized by its speakers.

A reality in which meaningful repetition is possible is a reality in which
finitary mathematics is possible.  Counting is built on the abstraction and
repetition of a meaningful property, like ``being an apple''.  An ``apple'' is a
``not-orange'', but it is also any apple.  Meaning is built on the
representation of repeatable aspects of (apparent) reality which do not
capture the totality of that reality.  An ``apple'' is a weight like this, and a
shape like this, and a smell like this, and a colour like this (or this, or this),
and a taste like this.  ``Like'' refers to a broad and imprecise range because,
in biological terms, it would be both impossible and counter-productive to
sense weights or shapes or smells or colours or tastes exactly.  We are able
to count apples, because it is possible, usually, for us to recognise an apple
when we see one, and to expect it to behave like a solid object as we move
it around.  An apple doesn't disappear without the possibility of an
explanation, such as that it has been eaten or stolen or hidden.  Finitary
mathematics tells us how collections of objects that tend not to disappear
must behave when the objects don't disappear.  For example,  $17 \times 19
= 18^2 -1$ because any collection of $324$ such objects can be moved from
a square array of side $18$ to a rectangular array of sides $17$ and $19$
with $1$ left over.  Marks on paper also tend not to disappear, so that we
can draw pictures and, eventually, develop a symbolism which allows us to
represent arithmetical laws which would apply to any finite collection of
stable objects.

The point I want to make here is not the dogmatic claim that finitary
mathematics is just a collection of facts about possible collections of
``objects'', but rather that stability and possibility and abstraction and
difference are utterly fundamental not only to the meaning of
mathematics, but also to the meaning of language, to the structure of our
minds, and to our ability to make sense of reality.  Mathematics is no more
just a language game than language is just a mathematics game.  I say
tomahto and you say tomayto, but what is essential is that neither of us
say grapefruit.  Language depends on abstractions and representations and
rules.  Language is possible if and only if finitary mathematics is possible. 
Mind too, characterized as a pattern of information defined by rules
representing elementary events which occur and change and recur, I
believe to be possible if and only if finitary mathematics is possible.

Possibility is fundamental to our reality, so it is not surprising that there are
many challenging metaphysical issues associated with it.  Among those
relevant to the present discusion are whether to be possible is to be;
how far possibility extends beyond ``reality''; and the relations
between possibility, truth, and fantasy, and between the finitary and the
infinitary.

In a philosophical analysis of the nature of possibility, Lewis
(1986) attempts a defense of the doctrine of modal realism -- the idea
that to be possible is to be.  Modal realism and mathematical realism have a
good deal in common.  A central question in both cases is how our thoughts
seem able to go beyond what we are, and the proposed answer in both
cases is that our thoughts can go beyond what we are because what there is
goes beyond what we are.  Both cases meet similar problems with the
infinite, with multiplicities of variant descriptions, and with the question of
how we can have knowledge of realities from which we are isolated.  Modal
realism has the additional problem that without some sort of a priori
measure on the space of possibilities, it seems hard to justify empirical
induction and to explain the apparent simplicity of our observed reality.  On
the other hand, I have a physical theory which allows me to define
precisely what I believe an observable possibility to be and how likely
each possibility is, but I do not believe that whether or not every one of
these possible experiences corresponds to an experienced reality makes
any difference to the experiences of those realities which are experienced.
This is just to say that, as with the problem of solipsism, no individual mind
in a ``many-minds'' theory can ever be certain, either from observing or
from infering the behaviour of other bodies, that other minds are real.
Indeed, because of this, leaving aside historical reasons and reasons to do
with the nature of quantum dynamics and how that dynamics is used to
calculate probabilities, it would make as much sense for me to refer to my
``many-minds'' interpretation of quantum theory as a ``possible-minds''
interpretation.

Related to the idea that possibilities cannot be understood unless they are
in some sense real, is the idea that the meaning of a number depends on
the actual physical existence of that many distinct entities, and so that 
there might be a largest actual finite number corresponding, in some way,
to the number of entities in the universe.  Even if I believed
the physics behind this claim, and of course I do not, it would not seem to
me that altering the actual size of the physical universe could make any
difference to the truth of any mathematical statement.  Nor do I accept
Rotman's (1993) arguments that we should limit our
imaginations by our apparent physical abilities.

The meanings of $(2816)^{71} + (2793)^{71} \ne (2834)^{71}$ include not
only facts about parity, but also facts about what would happen
if hyperbeings in $71 + 1$-dimensional spacetime tried re-arrangements
of hyperobjects arranged in hypercubes.  Given that we are not in
contact with them, whether there really are such beings is irrelevant to our
statements about what would happen if there were and to the meaning of
the formula for us.  Our minds always go beyond what we are
(Sartre 1943).  But they do not go beyond what we can imagine;
in other words, they do not go beyond the partial representations that we
can construct starting from our representations of what we are and what
we are not.  We are frustrated when our train is late, not because there is a
world somewhere in which the train is on time, but because we have seen
the timetable and know what we are entitled to expect.  Mathematics is an
extension of the framework of the constructions of our imagination. 
Mathematics builds into itself and makes explicit the laws of possibility --
the laws of logically consistent construction which allow us to distinguish
between possibility and impossibility.

All our lives we try to tame reality by telling stories.  Language, mind,
and mathematics are all part of this process.  For mind to be possible
there have to be stories to be told.  Mathematics is first and foremost about
mathematics, in the same way that ``The Lord of the Rings'' is first and
foremost about hobbits  -- we can tell stories which go way beyond
representations of our immediate biological needs and which build on their
own internal structures.

Fantasies are unrealistic stories.  To ride off into the setting sun is possible;
to live happily ever after is a fantasy.  The gulf between infinitary and
finitary mathematics is sufficiently wide that there is a sense of ``realistic''
according to which the processes of infinitary mathematics are fantasies,
but any finitary process in finitary mathematics is not.  To match a
rectangle of $(10^{436} - 1) \times (10^{436} + 1)$ physical objects with a
square of side $10^{436}$ less one would be possible if it were possible for
so many objects to be handled, but to match {\sl every\/} positive integer
with its double is a fantasy.  Finitary mathematics is concerned with
possibilities which are straightforward extensions of the possibilities within
which we live our mental lives.  There is no natural boundary within the
finite numbers at which any ``largest number'' could plausibly be located. 
$\infty$, however, does make a fundamental boundary.  I accept rules that
would govern arrangements of hyperobjects by hyperbeings in $71 +
1$-dimensional spacetime for the same reasons that I accept rules
governing arrangements of grains of rice on a table.  These rules depend on
logical consequence together with the idea of the predictable stability of
physical ``objects''.  The most important aspect of these finite possibilities is
that what would be involved in their achievement can be completely
understood.  After all, it seems about as likely that I will spend much time
arranging rice as that I will be contacted by hyperbeings, and I don't have
a horse.

As stories, fantasies can certainly be simple, coherent, and meaningful.  We
surely know what it means to match every positive integer with its
double.  ``Dot, dot, dot'' is an essential part of the abstraction which is
fundamental to our mental processes.  We all intend to live happily ever
after.  We can only realize our dreams one day at a time, but we rarely
know when the end will come.  If $n$ days are possible, then $n+1$ days
are possible, so, ignoring probability, we dream of forever.  Infinitary
mathematics is a fantasy world in which we fantasize about the
completions of processes which, realistically, we can only begin.  When we
write
$${\pi^2 \over 6} = \sum_{n=1}^\infty {1 \over n^2} = \sum_{n=1}^\infty
{1 \over (2n-1)^2} + \sum_{n=1}^\infty {1 \over (2n)^2} =
\sum_{n=1}^\infty {1 \over (2n-1)^2} + {\pi^2 \over 24}$$
and deduce that $\qquad \dsize \sum_{n=1}^\infty {1 \over
(2n-1)^2} = {\pi^2 \over 8}, \qquad$ we need some experience in the
manipulation of infinite series to be sure that what we are doing is justified.
That experience can be expressed by the $\varepsilon$ and $\delta$
arguments which reduce the manipulations to infinite classes of possible
manipulations involving finite sums.  We also need to justify the claim that
$\dsize{\pi^2 \over 6} - {\pi^2 \over 24} = {\pi^2 \over 8}$.  That
justification ultimately comes from experience with integers, with rationals,
with finite sequences, and with the logic of finite sentences and classes of
sentences such as (1).  Even if the laws of nature are
statements of infinitary mathematics, our understanding of those laws
remains founded on our finite experiences and it remains a fantasy to
imagine the processes of infinitary mathematics being realized by finite
beings. 

\proclaim{Conclusion.}
\endproclaim

Mathematics is about mathematics, but this is not to say that it is just a
social practice, or just a language game.  It is both, but it is also a means of
discovering truth.  It is useful to talk as if mathematical objects exist, just as
it is useful to talk as if characters in stories, or colours, or physical objects
exist.  The gulf between infinitary and finitary mathematics is
fundamental.  We are finite beings and the rules which structure our
minds, our languages, and our lives are the same as the rules which
underlie the truths of finitary mathematics.  We can always avoid the use
of infinitary mathematics in our scientific theories, but the cost may be the
simplicity of description which has led us to those theories.

As finite beings, the simplicity of infinitary mathematics is, for us, the
simplicity of fantasy.  As mathematicians, we build these fantasies; as
physicists, we try to grasp the fantasies which may actually govern our
reality; and as philosophers, we can fantasize about understanding it all.

\proclaim{Acknowledgements}{}
\endproclaim

I am grateful to Ed Segal and Ed Wallace for organizing a wide-ranging
seminar series and for giving me the opportunity to express the ideas in
this paper.  I am also grateful to David Corfield for many conversations
about the philosophy of mathematics.

\proclaim{References.}{}
\endproclaim

\frenchspacing
\parindent=0pt

\everypar={\hangindent=0.75cm \hangafter=1}

Bell, J.S. (1987)  {\sl Speakable and Unspeakable in Quantum
Mechanics.} (Cambridge)

Bratteli, O. and Robinson, D.W. (1981)  {\sl Operator Algebras and Quantum
Statistical Mechanics II.} (Springer)

Burgess, J.P. and Rosen, G. (1997)  {\sl A Subject with No Object.}
(Oxford)

Chihara, C.S. (1990)  {\sl Constructibility and Mathematical Existence.}
(Oxford)

DeWitt, B.S. and Graham, N. (1973)  {\sl The Many-Worlds Interpretation of
Quantum  Mechanics.}  (Princeton)

Donald, M.J. (1986)  ``On the relative entropy.''  {\sl Commun. Math. Phys.
}{\bf 105},  13--34.

Donald, M.J. (1992) ``A priori probability and localized observers.'' {\sl 
Foundations of Physics, \bf 22}, 1111--1172.

Donald, M.J. (1999)  ``Progress in a many-minds interpretation of
quantum theory.'' {\sl quant-ph/9904001}

Donald, M.J. (2002)  ``Neural unpredictability, the interpretation of
quantum theory, and the mind-body problem.''  {\sl  quant-ph/0208033}

{\hfill The core papers of my many-minds intepretation are also available
on my web site\hfill}

{\hfill {\catcode`\~=12 \catcode`\q=9
http://www.poco.phy.cam.ac.uk/q~mjd1014} \hfill}
\smallskip

Everett, H., III (1957) ``The theory of the universal wave function.'' 
In DeWitt and Graham (1973).

Haag, R. (1992)  {\sl Local Quantum Physics.} (Springer)

Halvorson, H. (2001)  ``On the nature of continuous physical quantities in
classical and quantum mechanics.''  {\sl Journal of Philosophical Logic
}{\bf 30},  27--50. 
{\sl quant-ph/0003074}

Hersh, R. (1997)  {\sl What is Mathematics, Really?}  (Random House)

Krieger, M.H. (1996) {\sl Constitutions of Matter.}  (Chicago)

Lewis, D. (1986)  {\sl On the Plurality of Worlds.} (Blackwell)

Maddy, P. (1997)  {\sl Naturalism in Mathematics.} (Oxford)

Quine, W.V. (1951)  ``Two dogmas of empiricism.''  {\sl Philosophical Review
}{\bf 60},  20--43.  (Often reprinted.)

Rotman, B. (1993)  {\sl Ad Infinitum.} (Stanford)

Ruelle, D. (1969)  {\sl Statistical Mechanics.} (Benjamin)

Sartre, J.-P. (1943) {\sl L'\^etre et le n\'eant.} (Gallimard)

Schr\"odinger, E. (1935)  ``Die gegenw\"artige Situation in der
Quantenmechanik.''  {\sl Naturwissenschaften} {\bf 23},  807--812,
823--828, 844--849.

Tasi\'c, V. (2001)  {\sl Mathematics and the Roots of Postmodern Thought.}
(Oxford)

\end